\documentclass[twocolumn,aps,prl]{revtex4}
\usepackage{graphicx}
\usepackage{bm}

\begin{document}
\title{Spin-polarized transport in ferromagnetic multilayered semiconductor nanostructures}
\author{E. J. R. Oliveira}
\affiliation{Instituto de F\'\i sica, Universidade do Estado do
Rio de Janeiro,  20.500-013 Rio de Janeiro, R.J., Brazil}
\author{A. T. da Cunha Lima}
\affiliation{Universidade Veiga de Almeida, Campus de Cabo Frio,
 28905-970 Cabo Frio, RJ, Brazil}
\author{M. A. Boselli }
\affiliation{Departamento de F\'\i sica, Universidade do Federal
de Ouro Preto, 35400-000 Ouro Preto, M.G., Brazil}
\author{G. M. Sipahi}
\affiliation{Instituto de F\'\i sica de S\~ao Carlos, Universidade
de S\~ao Paulo,  CP 369, 13560-970, S\~ao Carlos, SP, Brazil}
\author{S. C. P. Rodrigues}
\affiliation{Departamento de F\'{\i}sica, Universidade Federal de
Pernambuco, 50670-901 Recife, PE, Brazil }

\author{I. C. da Cunha Lima}
\affiliation{Instituto de F\'\i sica, Universidade do Estado do
Rio de Janeiro, 20.500-013 Rio de Janeiro, R.J.,
Brazil, and\\
Departamento de F\'\i sica, Universidade do Federal de Ouro Preto,
35400-000 Ouro Preto, M.G., Brazil}

\date{\today}

\begin{abstract}

The occurrence of inhomogeneous spin-density distribution in
multilayered ferromagnetic diluted magnetic semiconductor
nanostructures leads to strong dependence of the spin-polarized
transport properties on these systems. The spin-dependent
mobility, conductivity and resistivity in (Ga,Mn)As/GaAs,
(Ga,Mn)N/GaN, and (Si,Mn)/Si multilayers are calculated as a
function of temperature, scaled by the average magnetization of
the  diluted magnetic semiconductor layers. An increase of the
resistivity near the transition temperature is obtained. We
observed that the spin-polarized transport properties changes
strongly among the three materials.
\end{abstract}

\maketitle

Spin-polarized current and inhomogeneous spin density are two
elements to consider in Spintronics. During the last years the
fabrication and characterization of diluted magnetic semiconductor
(DMS) nanostructures has rapidly evolved. In the particular case
of DMS based on GaAs, GaN, ZnO, Si, and a few other semiconductors
a ferromagnetic phase exists with transition temperatures near and
even above room temperature \cite{matsu,dietl1}. Carriers in high
concentration  present in these systems play an important role in
the existence of the magnetic order. Ferromagnetism has been shown
to persist in some multilayered diluted magnetic heterostructures
(MDMH), structures with alternating non-magnetic and magnetic
layers \cite{bos,sado4} which produce an inhomogeneous spin
polarization density. In these MDMH the carriers come out of the
ionized magnetic atoms acting as donors or acceptors, even when
some compensation mechanism occur. In such structures the states
are spin-polarized and carriers with distinct spin-polarization
are scattered by the ionized atoms differently. Consequently, in
the ferromagnetic phase MDMHs show both spin-polarized currents
and inhomogeneous spin densities.

We present a calculation of the spin-polarized transport
properties of (Ga,Mn)As/GaAs, (Ga,Mn)N/GaN, and (Si,Mn)/Si MDMH as
a function of temperature. The aim of this paper is to explore the
interplay the presence of the magnetization and the
spin-dependent scattering processes as temperature changes.
We observe how conductivity, resistivity, and mobility
for each spin-polarization are affected as the sample average
magnetization changes with temperature. It has been shown
\cite{bos1} that an inhomogeneity in the spin density of
ferromagnetic MDMH structures is responsible for the high
transition temperatures observed.

We start with a model assuming a heterostructure in which: (i)
scattering centers occur only in certain layers, and they are the
very source of carriers; (ii) these layers are separated by
undoped spacers, ideally without any band mismatch; (iii) the
carriers are assumed to be in the metallic regime; (iv) barriers
represented by the potential $U_{\textrm{mod}}$ in the growth
direction are added to the layers, either to the spacers or to the
doped layers, having their height modulated by some external
parameter. Later we will relate them to the sample temperature.
Carriers are attracted to the doped layers by the Coulomb
interaction with the ionized impurities and are also affected by
the confining potential represented by the parametric barriers. In
the case of high carrier concentration a simplified model can be
used to obtain the electronic structure assuming that the
charged impurities are substituted by a homogeneous background, in
a kind of \textit{jellium} model.

The Hamiltonian of the motion in the $z$-direction reduces to that
of a free particle of effective mass $m^*$  plus terms due to the
confinement, $U_{\textrm{mod}}(z)$ and $V_C(z)$. The latter
incorporates the Hartree and the exchange-correlation terms of the
carrier-carrier interaction, and the interaction
carrier-background:
\begin{equation}
[-\frac{\hbar^2}{2m^*}\frac{\partial^2}{\partial
z^2}+U_{\textrm{mod}}(z)+V_C(z)]\Psi_n(z)=E_n\Psi_n(z).
\label{sch}
\end{equation}
The subbands are parabolic and have their energy given by
$E_n(k)=\hbar^2k^2/2m^* +E_n$, with $\textbf{k}=(k_x,k_y)$.
Eq.(\ref{sch}) and the Poisson equation are solved
self-consistently.

Next we calculate the in-plane transport relaxation time due to
scattering by ionized impurities based on the linearized Boltzmann
equation, as described in Ref.(\onlinecite{Bastard}). We neglect
inter-subband transitions. The current is obtained by summing up
the contribution of all occupied subbands, which behave as
different conducting channels. The mobility of each channel is
modulated by the barrier height, since it determines not only the
Fermi wavevectors of each subband, $k_F^{(n)}$, but also the
probability densities $|\Psi_n(z)|^2$, affecting directly the form
factor for an impurity located at $z=z_i$:
\begin{equation}
g_{\textrm{imp}}^{(n)}(\theta,z_i)=\int_{-\infty}^{\infty}
|\Psi_n(z)|^2\exp[-k_n(\theta)|z-z_i|]), \label{form}
\end{equation}
where $\theta$ is the scattering angle and we used
$k_n(\theta)=2k_F^{n}\sin \frac{\theta}{2}$. The inverse of the
momentum relaxation  time for subband $n$, using the screening
form factor $g_s(q)$ as in Ref. \onlinecite{Bastard}, results in:
\begin{eqnarray}
&&\frac{1}{\tau}_n=\frac{2e^2m^*n_i}{\hbar^3\kappa}\int_0^{\pi}d\theta
(1-\cos\theta)\times \nonumber \\
&&[k_n(\theta)+q_0g_s(k_n(\theta))]^{-2}\int dz
g(z)g_{\textrm{imp}}^{(n)}(\theta,z),\label{taun}
\end{eqnarray}
where $\kappa$ is the dielectric constant, $n_i$ the ionized
impurity concentration in each DMS layer, $q_0=n_s/a_0$, with
$n_s$ representing the effective areal carrier concentration,
$a_0$ the effective Bohr radius, and $g(z)$ equals 1 if $z$ lies
in the magnetic layer and zero otherwise.

\begin{figure}
\parbox{4cm}{
\includegraphics[angle=-90,width=4.5cm,bb = 15 -20 597 732]{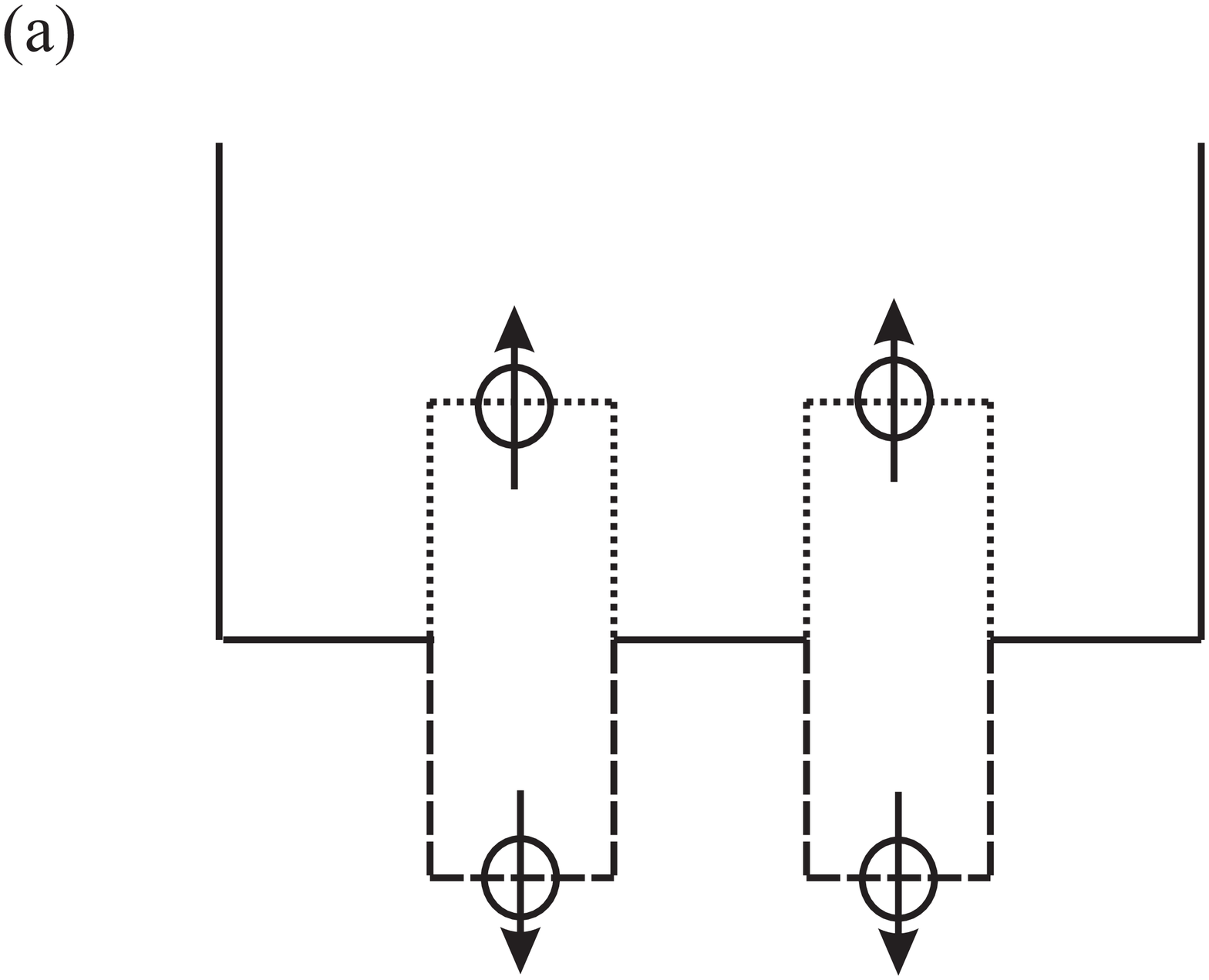}
} {\hspace{\fill}
\includegraphics[angle=-90,width=4.2cm, bb= 315 91 582 782]{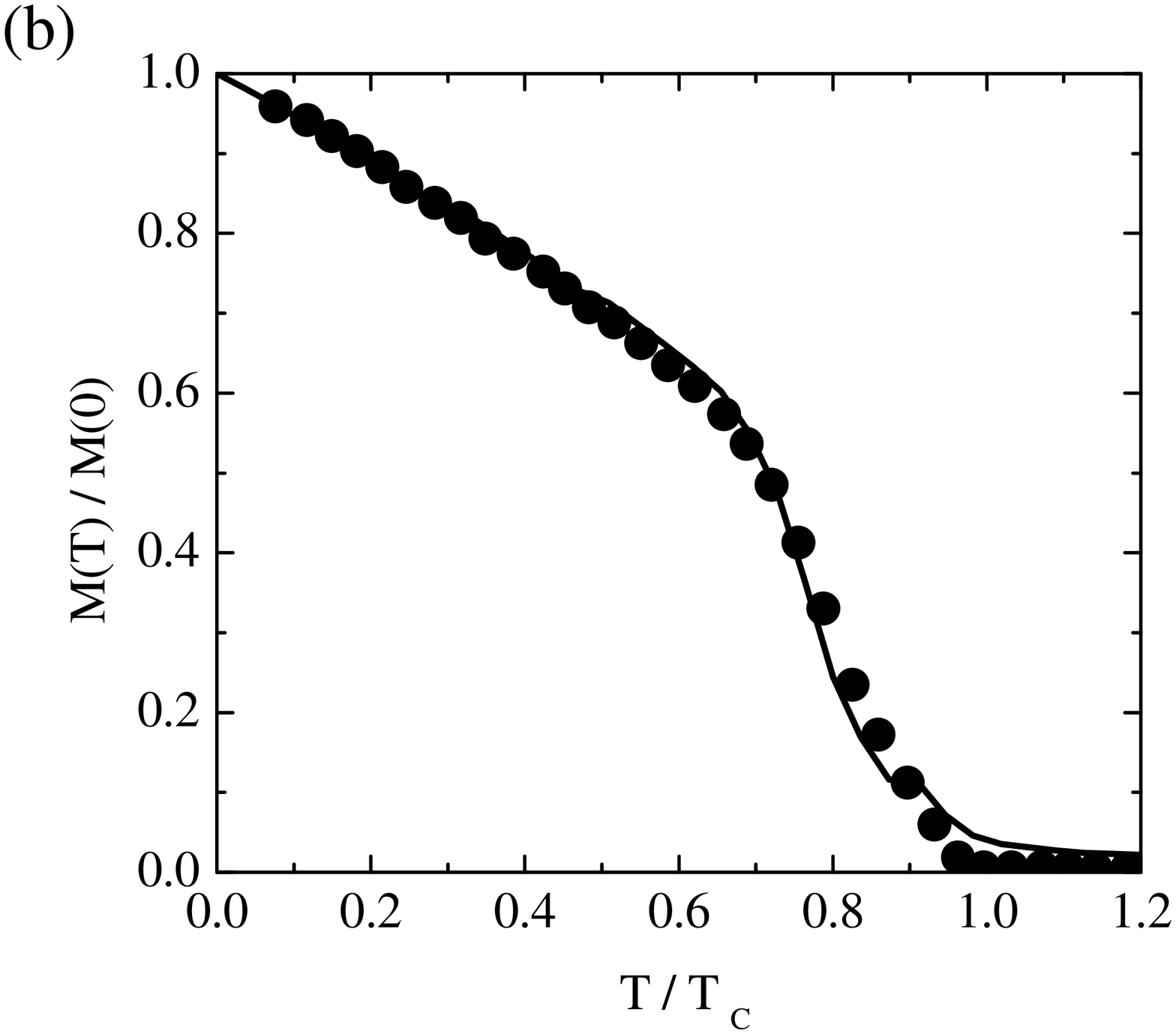}
} \caption{(a)Effective magnetic barriers for holes with spins
aligned (up) and anti-aligned (down) to the average magnetization.
(b)Modulation function for the potential barrier  extracted from
the curve of the magnetization  versus temperature. Solid line
after Ref.\onlinecite{bos}, dots after Ref.\onlinecite{sado4}.}
\label{fig1}
\end{figure}

The in-plane transport of MDMHs can be mapped into this model. In
mean field approximation an average magnetization $<M>$ of the
magnetic layers provides a potential barrier which is $\pm$75 meV
for Ga$_{0.95}$Mn$_{0.05}$As at T=0 K, as shown in Fig.
\ref{fig1}a. Free carriers are heavy holes. It has been observed
that in MDMH structures the magnetization changes with temperature
according to the curve shown in Fig. \ref{fig1}b. This curve is
used here as a scale relating the temperature to the magnetic
barriers. This Ansatz provides the dependence of the
spin-dependent modulated potential $U_{\textrm{mod}}(z)$ in
Eq.(\ref{sch}) with the sample temperature. It contains the
complete information about the magnetic part of the system.  On
the other hand, the calculation of the transport relaxation time
depends on the concentration of ionized impurities, which is taken
as equal to the carrier concentration. The limit $T=0$ is assumed
for the equilibrium distribution in the linearized Boltzmann
equation, but a finite $T$ is taken as a parameter modulating the
barrier height. In this sense, we define a normalized conductivity
$\sigma(T)/\sigma(0)$ as the ratio between the conductivities
resulting of the average magnetization at temperature $T$, and the
maximum magnetization obtained when $T=0$. We can also obtain the
fraction of the normalized conductivity coming out of spins
aligned (up) or anti-aligned (down) to the average magnetization
by summing up the contributions of the channels with equal spin
polarization.

\begin{figure}[h]

\includegraphics[angle=-90,width=\columnwidth]{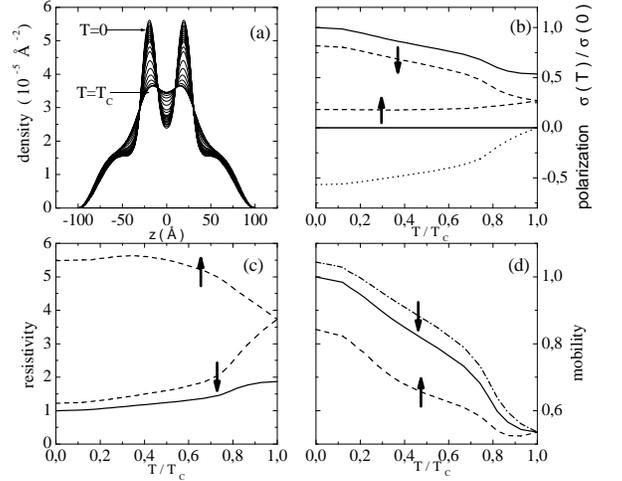}
\vspace*{-1cm}\caption{(Ga,Mn)As/GaAs MDMH: (a) charge density
distribution; (b) normalized conductivities and spin-
polarization; (c) spin-polarized and total resistivities; (d)
spin-dependent and total mobility. Total quantities are indicated
by full lines, spin-dependent quantities by dashed lines with
arrows indicating parallel ($\uparrow$) or anti-parallel
($\downarrow$) polarization.} \label{fig2}
\end{figure}

The calculations were performed for (Ga,Mn)As/GaAs, (Ga,Mn)N/GaN,
and (Si,Mn)/Si. In each case we calculated (a) the areal charge
densities as a function of the position in the z-axis for
different temperatures; (b) the normalized spin-dependent and
total conductivity together with the polarization; (c) the
spin-polarized and total resistivities; (d) the spin-polarized
and total mobilities.

\begin{figure}
\includegraphics[angle=-90,width=\columnwidth]{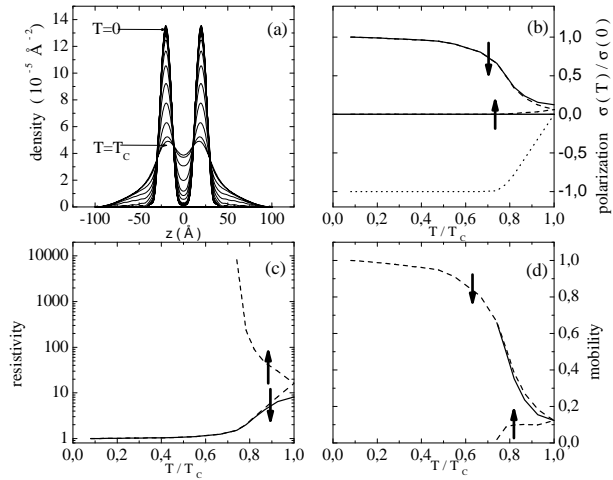}
\vspace*{-1cm}\caption{For (Ga,Mn)N/GaN, as above, using a
logarithmic scale for the resistivity.}\label{fig3}
\end{figure}

\begin{figure}[h]
\includegraphics[angle=-90,width=\columnwidth]{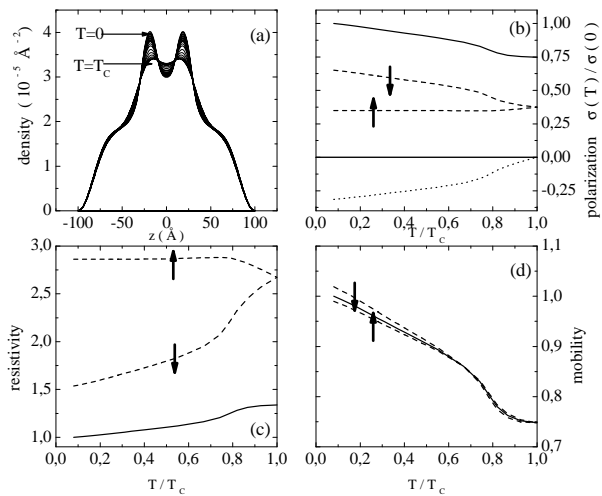}
\vspace*{-1cm}\caption{Same as above for (Si,Mn)/Si}\label{fig4}
\end{figure}

The carrier concentration inside the DMS layers decreases as the
temperature increases, showing that the magnetic barriers compete
with the Coulomb attraction due to ionized atoms at low
temperature, but loses importance as the temperature increases, as
expected. Observing Figs. \ref{fig2}(a), \ref{fig3}(a) and
\ref{fig4}(a) we see that, in the case of (Ga,Mn)N/GaN the
concentration of the charge density inside the DMS layers is
almost complete, even at temperatures near $T_C$. For (Si,Mn)/Si
the concentration is the weakest among the three systems. As a
consequence of the strength of the magnetic barriers, carriers in
(Ga,Mn)/GaN are completely spin-polarized even close to $T_C$. The
polarization is weaker for (Ga,Mn)As/GaAs, and weaker still for
(Si,Mn)Si.

The conductivity for each spin-dependent channel depends not only
on the transport relaxation time, but also on the density of
carriers occupying the subband. We show in Figs.
\ref{fig2}(b)-\ref{fig4}(b)  the normalized conductivity, defined
as $\sigma(T)/\sigma(T_C)$. For (Ga,Mn)As/GaAs and (Ga,Mn)N/GaN,
there is a remarkable difference between the conductivity for
channels with spins up and down.  For (Si,Mn)/Si, with full
spin-polarization, the conductivity is zero for spins up until
very near $T_C$. The resistivities, shown in
Figs.\ref{fig2}(c)-\ref{fig4}(c), reflect the behavior of the
conductivity. It is worthwhile to stress  the fact that a hump
appears near $T_C$, corresponding to the increase of the
resistivity  as the transition temperature is aproached. This
behavior was observed in (Ga,Mn)As epilayers \cite{matsu}, and is
a consequence of the change in the subbands population as the
magnetic barriers decrease drastically when the transition
temperature is reached. Finally, it is important to obtain the
mobility, since it is directly related to the average velocity of
carriers with a given spin-polarization. While the mobility in
(Si,Mn)/Si is practically spin-independent, in (Ga,Mn)As/GaAs the
difference is almost 30\% for spins up and down.  For (Ga,Mn)N/GaN
we observe a kind of half-metallic behavior, where carriers with
spins up have very small mobility until very near to $T_C$, while
carriers with spin down are metallic.

Besides the fact that our calculation explains the hump in the
resistivity near $T_C$, we have shown that the charge density and
spin density inhomogeneities resulting of the effective magnetic
barriers in MDMH can be used to provide highly spin-polarized
resistivity and mobility. Certainly, spin-flip mechanisms as well
as interaction with phonons must be included if a more complete
treatment near the transition temperature is desired. However, we
believe that MDMH are useful at a temperature range far enough
from $T_C$ to make our calculation realistic.

This work was partially supported by CNPq (ESN fellowship and
research grant), FAPEMIG, FAPESP and FAPERJ. ICCL is grateful for
the hospitality of Prof. M. W. Wu group at the USTC, Hefei, Anhui,
China.

\end{document}